\documentclass[sigconf, natbib=true, anonymous=false]{acmart}
\makeatletter
\def\@ACM@checkaffil{
    \if@ACM@instpresent\else
    \ClassWarningNoLine{\@classname}{No institution present for an affiliation}%
    \fi
    \if@ACM@citypresent\else
    \ClassWarningNoLine{\@classname}{No city present for an affiliation}%
    \fi
    \if@ACM@countrypresent\else
        \ClassWarningNoLine{\@classname}{No country present for an affiliation}%
    \fi
}
\makeatother

\usepackage{graphicx}
\usepackage{amsmath}
\usepackage{enumitem}
\usepackage{multirow}
\usepackage{booktabs}
\usepackage{tabularx}
\usepackage[normalem]{ulem}
\useunder{\uline}{\ul}{}

\setlist[description]{
    labelwidth=0pt,
    leftmargin=0pt,
    itemindent=\labelsep\relax,
}
\begin{document}

\title{InRanker: Distilled Rankers for Zero-shot Information Retrieval}

\author{Thiago Laitz}
 \affiliation{%
   \institution{UNICAMP, FEEC, Brazil}
 }

\author{Konstantinos Papakostas}
\affiliation{%
   \institution{Zeta Alpha, Netherlands}
 }

\author{Roberto Lotufo}
\affiliation{%
   \institution{UNICAMP, FEEC, Brazil}
 }
\affiliation{%
   \institution{NeuralMind, Brazil}
 }
 
\author{Rodrigo Nogueira}
\affiliation{%
   \institution{UNICAMP, FEEC, Brazil}
 }
\affiliation{%
   \institution{Zeta Alpha, Netherlands}
 }


\begin{abstract}
    Despite multi-billion parameter neural rankers being common components of state-of-the-art information retrieval pipelines, they are rarely used in production due to the enormous amount of compute required for inference. In this work, we propose a new method for distilling large rankers into their smaller versions focusing on out-of-domain effectiveness. We introduce InRanker, a version of monoT5 distilled from monoT5-3B with increased effectiveness on out-of-domain scenarios. Our key insight is to use language models and rerankers to generate as much as possible synthetic "in-domain" training data, i.e., data that closely resembles the data that will be seen at retrieval time. The pipeline consists of two distillation phases that do not require additional user queries or manual annotations: (1) training on existing supervised soft teacher labels, and (2) training on teacher soft labels for synthetic queries generated using a large language model. Consequently, models like monoT5-60M and monoT5-220M improved their effectiveness by using the teacher's knowledge, despite being 50x and 13x smaller, respectively. Models and code are available at \url{https://github.com/unicamp-dl/InRanker}
    
\end{abstract}

\maketitle

\section{Introduction}
It is well known that the effectiveness of IR pipelines increases with larger models~\cite{bonifacio2022inpars,neelakantan2022text,ni2021large,nogueira2020document,pradeep2021expando}. For instance, multi-billion parameter rankers and dense models achieve top positions on leaderboards of IR benchmarks and competitions~\cite{trec2020,trec2021,trec2022}. These large models leverage increased representation capacity, enabling them to encode features that might elude smaller models. However, deploying these large models is not without its challenges. The computational overheads are substantial, often requiring specialized hardware such as GPUs or TPUs to operate in latency-critical applications. The high cost is directly related to the large number of parameters that these models contain, as they require hardware with high memory and compute capacity. In a production environment, this means higher operating costs and reduced scalability.

To address these challenges, there have been efforts to create more efficient models without significantly reducing effectiveness. One such approach is model distillation~\cite{hinton2015distilling}. Distilled models, such as MiniLM~\cite{minilm}, use a teacher or an ensemble of larger models to transfer knowledge to a smaller student model. \citet{rosa2022parameter} show that MiniLM surpassed the zero-shot effectiveness of monoT5-base in IR tasks despite being an order of magnitude smaller in size. This has shown that knowledge transfer via model distillation is not only feasible but also effective. However, most distillation techniques have been geared towards optimizing effectiveness on specific benchmark tasks and do not focus on out-of-domain effectiveness. Rosa et al. also show that while smaller models are capable of achieving high in-domain results, similar to their larger counterparts, the disparity in effectiveness becomes evident in out-of-domain scenarios. 

    \begin{figure}[t]
    \centering
    \includegraphics[trim={1cm 0cm 1cm 1cm}, clip, width=1\linewidth]{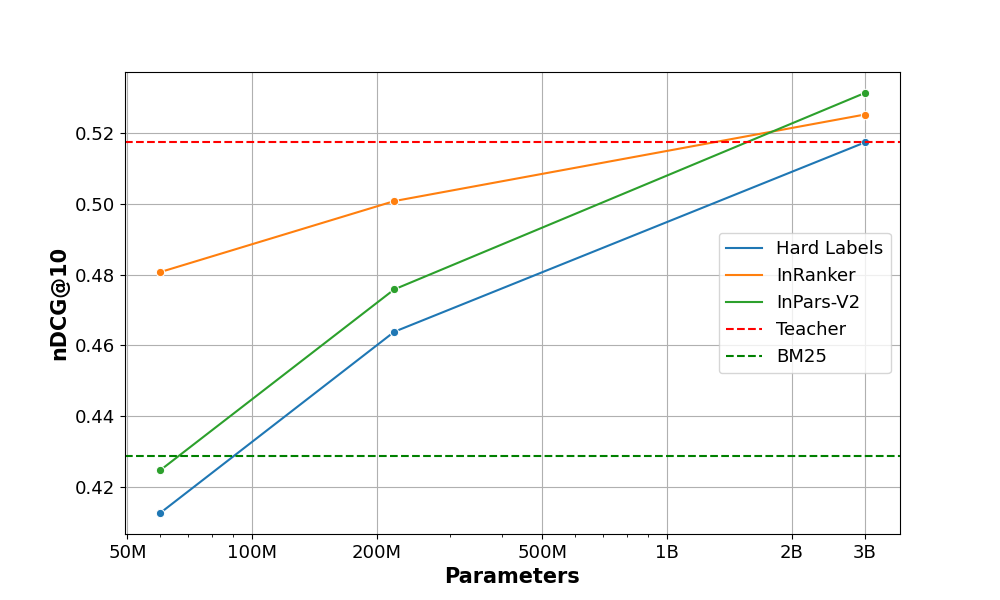}
    \caption{Effectiveness on the BEIR benchmark~\cite{beir}. All models are based on monoT5~\cite{nogueira2020document}, applying different finetuning methods.}
    \label{fig:plot_ndcg}
\end{figure}

Usually, training a retrieval model requires human-annotated hard labels informing which passage is relevant for each query. However, with the advance of Large Language Models (LLMs), it has become possible to generate synthetic queries for passages, providing a feasible approach for data augmentation~\cite{bonifacio2022inpars,jeronymo2023inpars,boytsov2023inpars,10.1145/3543507.3583261,10.1145/3539618.3591960}. Our work introduces a method for the generation of synthetic data specifically designed for distilling rankers that increases their out-of-domain effectiveness. We present InRanker, a distilled model derived from monoT5-3B~\cite{nogueira2020document}, that uses the predictions of the teacher directly with both synthetic, generated from an out-of-domain corpus, and real query-document pairs. Effectively, this approach converts any corpus to be in-domain, since the model will be trained using queries from the target domain. As a result, this approach leads to reduced model sizes while maintaining improved out-of-domain effectiveness as presented in Figure~\ref{fig:plot_ndcg}. The methodology, results, and ablation experiments are presented in detail in the following sections.

 \begin{figure*}[t]
    \centering
    \includegraphics[width=0.85\linewidth]{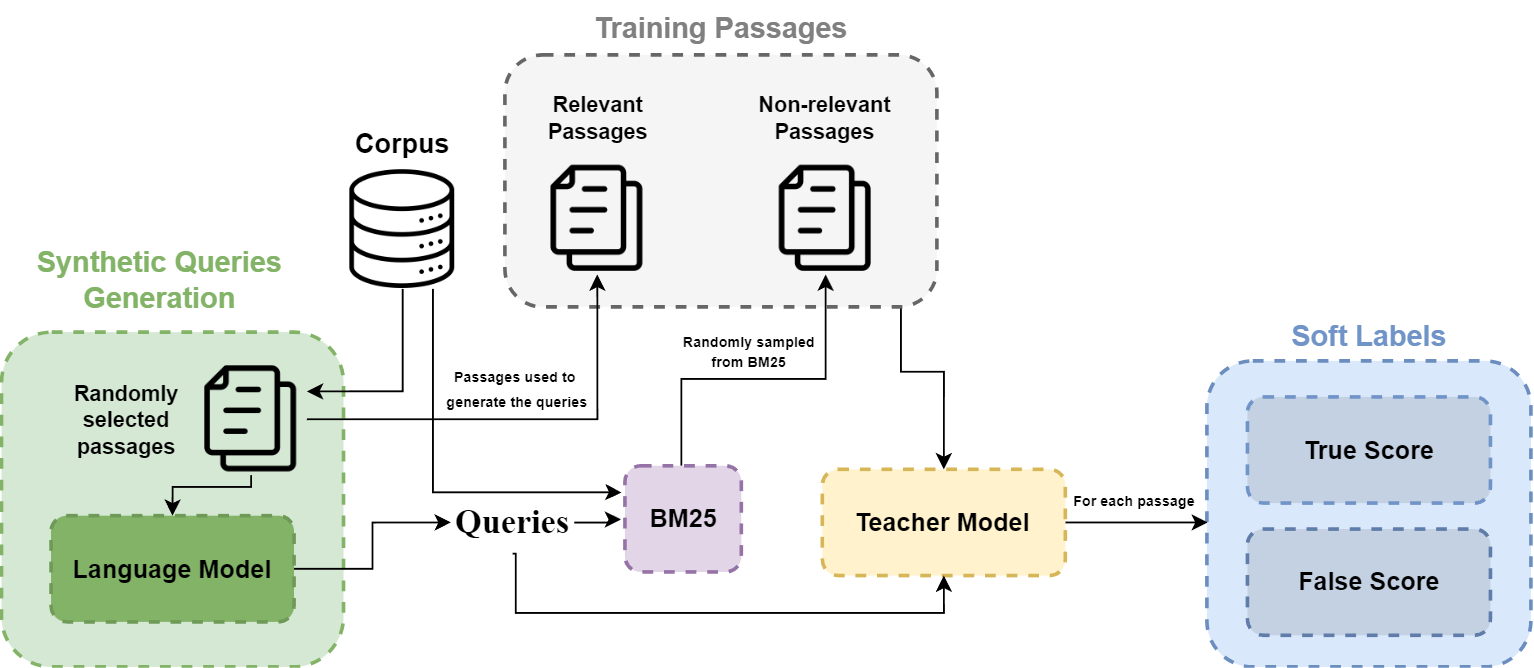}
    \vspace{0.1cm}
    \caption{Pipeline for generating the synthetic triples <query, passage, soft label> for the InRanker model.}
    \label{fig:phase1}
\end{figure*}

\section{Related Work}
The research community has been using LLMs in a variety of tasks aimed at increasing the availability of data and improving the effectiveness of existing systems. \citet{magister2023teaching} employed synthetic text generated by PaLM 540B~\cite{chowdhery2022palm} and GPT-3 175B~\cite{brown2020language} to transfer knowledge to smaller models such as T5. \citet{fu2023specializing} successfully specialized student models in multi-step reasoning using FlanT5~\cite{flant5} and code-davinci-002 as teachers. However, all these works rely on training the student models using synthetic text rather than directly using the soft labels. Furthermore, \citet{muhamed2021ctr} distilled cross-attention scores of a language model for click-through-rate prediction, achieving better results when exposed to contextual features such as tabular data. \citet{minilm} distilled the self-attention module, which is a crucial part of transformers, and successfully transferred knowledge to a variety of tasks.

Previous studies have also explored training a student from soft labels produced by a teacher: \citet{hofstätter2021improving} proposed a cross-architecture knowledge distillation approach using the MarginMSE loss. Similarly, \citet{formal2022distillation} used the MarginMSE loss to distill knowledge for sparse neural models. Finally, \citet{hashemi2023dense} proposed a method for generating synthetic data for domain adaptation of dense passage retrievers. This approach involves creating new queries and a target collection, along with pseudo-labels extracted using a BERT cross-encoder. However, they did not evaluate the model's effectiveness on datasets to which it was not domain-adapted. The existing research has mainly focused on in-domain evaluation, where the goal has been to increase the effectiveness of the student model on test datasets whose domain is similar to the datasets it was trained on. Our study also focuses on the robustness of the student and its ability to perform well even in out-of-domain scenarios, similar to the abilities of the larger teacher model.
    
\section{Methodology}
Our proposed method consists of two key phases of distillation, each designed with specific objectives to maximize the model's zero-shot effectiveness. The first phase uses real-world data to familiarize the student model with the ranking task, while the second phase uses synthetic data designed to improve zero-shot generalization and improve the model's effectiveness on a specific dataset. The dataset used to distill InRanker consists of \{query, passage, logits\} triplets, where the logits (soft labels) originate from a teacher model that has been trained for the relevance task.
For the first stage, we chose to use query-document pairs from the MS MARCO~\cite{msmarco} dataset, given their variety, the large number of annotated pairs, and its demonstrated effectiveness in enhancing retrieval effectiveness~\cite{ren2023examination}. Next, we source the synthetic queries from InPars~\cite{bonifacio2022inpars}, which used an LLM to generate queries for the datasets in BEIR in a few-shot manner.

Distilling rerankers involves using the Mean Squared Error (MSE) loss to match the logits of the teacher and the student, as part of a two-phase pipeline illustrated in Figure~\ref{fig:phase1}. The first phase consists of two steps: (1) generating the teacher logits given a query and either a positive (relevant) or a negative (non-relevant) passage, where the negatives are randomly sampled using BM25 on the top-$k=1000$ candidates, and the positives are sampled from the human-annotated pairs; and (2) training InRanker given the queries and passages as input using the MSE loss to match the student logits to those of the teacher, who remains frozen during training. This approach can be beneficial as it removes the need for making hard decisions about a passage's relevance, i.e. determining a threshold to obtain binary relevance labels, and instead focuses on a soft target objective aimed at aligning the student's perception of relevance with that of the teacher.
    
The second phase, with a focus on zero-shot effectiveness, uses the same two steps. However, instead of employing real queries sourced from a costly human-annotation process, it utilizes synthetic queries generated by an LLM based on randomly sampled documents from the corpus. In this scenario, the positive document is the one used to create the query, and the negatives are collected using the same top-$k$ sampling approach as before.

We also perform zero-mean normalization on the teacher logits for each query-document pair, independent of the overall dataset distribution. This approach intends to make the data distribution symmetric for each query-document pair, thereby minimizing the bias that InRanker is required to learn. Formally:
\begin{align}
    \begin{split}
        {L'}_\mathrm{true} &= L_\mathrm{true} - \frac{L_\mathrm{true} + L_\mathrm{false}}{2} \\
        {L'}_\mathrm{false} &= L_\mathrm{false} - \frac{L_\mathrm{true} + L_\mathrm{false}}{2}
    \end{split}
    \label{eq:normalization}
\end{align}
with $L_\mathrm{true}$ and $L_\mathrm{false}$ denoting the teacher's logits for the relevant and non-relevant classes, respectively, and $L'$ being the normalized values.
This results in the following loss for each training example:
\begin{equation}
    \mathcal{L}_\mathrm{MSE} = ([Y_\mathrm{true} - {L'}_\mathrm{true}]^2 + [Y_\mathrm{false} - {L'}_\mathrm{false}]^2)
    \label{eq:mse}
\end{equation}
with $Y_\mathrm{true}$ and $Y_\mathrm{false}$ representing the logits of the student.

Due to the training objective described in equation \eqref{eq:mse}, the model no longer determines the relevance of passages and instead focuses on replicating the teacher's output, thus eliminating the need for tunning a relevance threshold that would be needed to produce a binary label. With this approach, we can easily expand the out-of-domain knowledge of distilled models by generating new queries for documents using an LLM and fine-tuning the distilled model using the teacher's logits. In the experiments section, we demonstrate the effectiveness of this approach in enhancing the student model's effectiveness across 16 datasets of BEIR simultaneously. We present the hyperparameters used for training and the dataset curation in Appendix~\ref{sec:training}, and we discuss variations of the training loss in Appendix~\ref{sec:losses}.

\section{Experiments}

\subsection{Knowledge Distillation Results}
We distilled \href{https://huggingface.co/castorini/monot5-3b-msmarco-10k}{monoT5-3B} to models with parameters ranging from 60M to 3B, using combinations of the following configurations:
\begin{description}
    \item[Human Hard:] representing the common approach for training rankers with human-annotated hard (i.e., binary) labels from the MS MARCO passage ranking dataset. In this case, a vanilla cross-entropy loss is used:
    \begin{equation}
    \mathcal{L}_\mathrm{CE} = -\log P_\mathrm{relevant} - \log P_\mathrm{non-relevant}
    \label{eq:ce}
\end{equation}
    where $P_\mathrm{relevant}$ and $P_\mathrm{non-relevant}$ are the probabilities assigned by the model to the relevant and non-relevant query-document pair, respectively. Non-relevant pairs are sampled from the top-1000 retrieved by BM25.
    \item[Human Soft:] representing a distillation step for matching the logits of a teacher and a student model, using real (human-generated) queries from the ranking dataset as inputs, but without the binary relevance judgments for targets.
    \item[Synthetic Soft:] representing a distillation step for matching the logits of the two models, similar to the previous configuration, but using exclusively synthetic queries generated from the corresponding BEIR corpora with InPars~\cite{bonifacio2022inpars, jeronymo2023inpars}.
\end{description}

\begin{table}[ht]
    \centering
    \renewcommand{\arraystretch}{1.1}
    \setlength{\tabcolsep}{5pt}
    \resizebox{\columnwidth}{!}{%
      \begin{tabularx}{\columnwidth}{@{}lcccc@{}}
        \toprule
         & \multicolumn{3}{c}{\textbf{Training Configurations}} & \\ \cmidrule(lr){2-4}
        \multicolumn{1}{c}{{\textbf{Model}}} & \begin{tabular}[c]{@{}c@{}}Human\\ Hard\end{tabular} & \begin{tabular}[c]{@{}c@{}}Human\\ Soft\end{tabular} & \begin{tabular}[c]{@{}c@{}}Synthetic\\ Soft\end{tabular} & \textbf{\begin{tabular}[c]{@{}c@{}}Avg.\\Score\end{tabular}} \\ \midrule
        (1) monoT5-60M & $\checkmark$ &  &  & 0.4125 \\
        (2) $\hookrightarrow$ w/ soft human &  & $\checkmark$ &  & 0.4356 \\
        (3) InRanker-60M &  & $\checkmark$ & $\checkmark$ & \textbf{0.4807} \\ \midrule
        (4) monoT5-220M & $\checkmark$ &  &  & 0.4638 \\
        (5) $\hookrightarrow$ w/ soft human &  & $\checkmark$ &  & 0.4870 \\
        (6) InRanker-220M &  & $\checkmark$ & $\checkmark$ & \textbf{0.5008} \\ \midrule
        (7) monoT5-3B* & $\checkmark$ &  &  & 0.5174 \\
        (8) InRanker-3B & $\checkmark$ &  & $\checkmark$ & \textbf{0.5253} \\ \bottomrule
  \end{tabularx}
    }
    \vspace{0.1cm}
    \caption{Distillation results (nDCG@10) on 16 BEIR datasets. The model marked with * represents the teacher model. We did not train InRanker-3B on human soft labels due to computational constraints.}
    \label{table:results}
\end{table}

From Table \ref{table:results}, we see that both distillation steps were essential for improving the average nDCG@10 score compared to the model trained solely using human hard labels from MS MARCO.\footnote{Results per dataset are shown in Appendix~\ref{appendix:full_results}.} As a result, InRanker-60M (row 3) and InRanker-220M (row 6), despite being 50x and 13x smaller than the teacher model, were able to improve their effectiveness on the BEIR benchmark significantly. Moreover, models trained exclusively on MS MARCO soft labels (rows 2 \& 5) saw an increase in effectiveness in comparison to training on solely hard labels (rows 1 \& 4), corroborating findings from previous studies regarding the effectiveness of soft labels \cite{hinton2015distilling,hofstätter2021improving,formal2022distillation,hashemi2023dense}. Furthermore, we observed an increase in the effectiveness even in self-distillation training (row 8), where the student learns soft labels generated by itself. We hypothesize that the improvement stems from the extra knowledge provided by the language model used to generate the synthetic queries. We did not provide results for the 3B model trained on both human soft and synthetic soft due to computational costs.

Furthermore, in Table~\ref{table:comparison}, we present a effectiveness comparison between InRanker, Promptagator~\cite{promptagator}, and RankT5~\cite{rankt5}. Although we used monoT5-3B as a teacher for our experiments, which has a lower effectiveness on average when compared to Promptagator or RankT5-3B, our method is model-agnostic and thus one could use a stronger teacher model and anticipate even stronger results. Nonetheless, InRanker remains competitive in both model groups of 220M and 3B parameters, outperforming the other two baselines in 6 out of the 10 evaluated datasets, despite the average score not reflecting this due to Promptagator and RankT5 attaining a significantly higher score in two datasets: ArguAna and Touché.

\begin{table*}[ht]
    \centering
    \renewcommand{\arraystretch}{1.2}
    \resizebox{.9\textwidth}{!}{%
    \begin{tabular}{@{}lc|ccc|ccc@{}}
        \toprule
        \multicolumn{1}{c}{\textbf{Dataset}} & \textbf{\begin{tabular}[c]{@{}c@{}}InRanker\\ 60M\end{tabular}} & \textbf{\begin{tabular}[c]{@{}c@{}}InRanker\\ 220M\end{tabular}} & \textbf{\begin{tabular}[c]{@{}c@{}}Promptagator++\\ 110M + 110M\end{tabular}} & \textbf{\begin{tabular}[c]{@{}c@{}}RankT5-Enc\\ 220M\end{tabular}} & \textbf{\begin{tabular}[c]{@{}c@{}}InRanker\\ 3B\end{tabular}} & \textbf{\begin{tabular}[c]{@{}c@{}}monoT5\\ 3B*\end{tabular}} & \textbf{\begin{tabular}[c]{@{}c@{}}RankT5-Enc\\ 3B\end{tabular}} \\ \hline
        TREC-COVID & \multicolumn{1}{c|}{0.7775} & \textbf{0.7984} & 0.7620 & 0.7896 & 0.8175 & 0.7936 & {\ul \textbf{0.8237}} \\
        NFCorpus & \multicolumn{1}{c|}{0.3547} & 0.3658 & 0.3700 & \textbf{0.3731} & 0.3825 & 0.3801 & {\ul \textbf{0.3990}} \\
        HotpotQA & \multicolumn{1}{c|}{0.7563} & \textbf{0.7742} & 0.7360 & 0.7269 & {\ul \textbf{0.7800}} & 0.7595 & 0.7536 \\
        Climate-FEVER & \multicolumn{1}{c|}{0.2729} & \textbf{0.2914} & 0.2030 & 0.2462 & {\ul \textbf{0.2931}} & 0.2835 & 0.2753 \\
        DBPedia & \multicolumn{1}{c|}{0.4451} & \textbf{0.4650} & 0.4340 & 0.4373 & {\ul \textbf{0.4762}} & 0.4719 & 0.4598 \\
        ArguAna & \multicolumn{1}{c|}{0.2466} & 0.2873 & {\ul \textbf{0.6300}} & 0.3094 & \textbf{0.4243} & 0.3824 & 0.4069 \\
        Touché-2020 & \multicolumn{1}{c|}{0.2883} & 0.2897 & 0.3810 & \textbf{0.4449} & 0.2924 & 0.3026 & {\ul \textbf{0.4869}} \\
        SCIDOCS & \multicolumn{1}{c|}{0.1788} & 0.1911 & {\ul \textbf{0.2010}} & 0.1760 & \textbf{0.1990} & 0.1978 & 0.1918 \\
        SciFact & \multicolumn{1}{c|}{0.7490} & \textbf{0.7618} & 0.7310 & 0.7493 & {\ul \textbf{0.7831}} & 0.7773 & 0.7600 \\
        FiQA-2018 & \multicolumn{1}{c|}{0.4043} & 0.4431 & \textbf{0.4940} & 0.4132 & 0.5027 & {\ul \textbf{0.5068}} & 0.4932 \\ \hline
        Average & \multicolumn{1}{c|}{0.4474} & 0.4668 & \textbf{0.4942} & 0.4666 & 0.4951 & 0.4856 & {\ul \textbf{0.5050}} \\ \bottomrule
    \end{tabular}%
    }
    \vspace{0.1cm}
    \caption{Comparison of the effectiveness for various reranking models, measured by nDCG@10 on the BEIR benchmark. The model marked with * represents the teacher model used for training InRanker. Bolded scores correspond to the best effectiveness on a specific dataset for a given model size, while underlined scores indicate the best effectiveness overall.}
    \label{table:comparison}
\end{table*}

\subsection{Ablations}

In this section, we present our ablation experiments aimed at validating the best configuration for distilling monoT5-3B into smaller T5-based models, as well as assessing their zero-shot capabilities. The initial experiments we conducted focused on evaluating how distillation would affect the model's effectiveness on novel dataset distributions that were not seen during training, i.e., we did not generate synthetic queries for them. To achieve this, we created two subsets, each containing $8$ randomly selected datasets from 16 datasets of BEIR\footnote{We list their exact composition in Appendix~\ref{sec:datasets}.}, which we named sample sets 1 and 2 and used only one set for training per experiment. The datasets that were used for training are designated as the ``in-domain'' category, while the remaining datasets, i.e. the other 8 datasets that are not part of the training set, represent the ``out-of-domain'' (O.O.D.) category. 

\begin{table*}[tb]
\centering
\renewcommand{\arraystretch}{1.1}
{%
    \begin{tabularx}{.8\textwidth}{@{}l *8{>{\centering\arraybackslash}X}@{}}
        \toprule
         &
        \multicolumn{3}{c}{\textbf{Training Configurations}} &
        \multicolumn{2}{c}{\textbf{Sample Set 1}} &
        \multicolumn{2}{c}{\textbf{Sample Set 2}} \\ \cmidrule(lr){2-4} \cmidrule(lr){5-6} \cmidrule(lr){7-8}
        \multicolumn{1}{c}{{\textbf{T5 Model}}}  & Human Hard & Human Soft & Synthetic Soft & In-domain & O.O.D. & In-domain & O.O.D. \\ \midrule
        (1) 60M (monoT5) & $\checkmark$ & & & 0.4141 & 0.4109 & 0.4817 & 0.3434 \\
        (2) 60M & & $\checkmark$ & & 0.4422 & 0.4290 & 0.5124 & 0.3587 \\
        (3) 60M (InRanker) & & $\checkmark$ & $\checkmark$ & \textbf{0.4768} & \textbf{0.4716} & \textbf{0.5558} & \textbf{0.3852} \\
        (4) 60M & $\checkmark$ & & $\checkmark$ & 0.4475 & 0.4587 & 0.5355 & 0.3617 \\ \midrule
        (5) 220M (monoT5) & $\checkmark$ & & & 0.4647 & 0.4629 & 0.5475 & 0.3801 \\
        (6) 220M & & $\checkmark$ & & 0.4867 & 0.4873 & 0.5692 & 0.4048 \\
        (7) 220M (InRanker) & & $\checkmark$ & $\checkmark$ & \textbf{0.4945} & \textbf{0.5028} & \textbf{0.5874} & \textbf{0.4083} \\
        (8) 220M & $\checkmark$ & & $\checkmark$ & 0.4905 & 0.4942 & 0.5832 & 0.3941 \\ \midrule
        (9) 3B* (monoT5) & $\checkmark$ & & & 0.5095 & 0.5253 & 0.6053 & 0.4295 \\ \bottomrule
    \end{tabularx}
}
\vspace{0.1cm}
\caption{Comparison of the in-domain vs out-of-domain effectiveness of our method, measured by nDCG@10. The model marked with * represents the teacher model used for the knowledge distillation process.}
\label{table:ablation}
\end{table*}

\paragraph{\textbf{Impact of soft knowledge distillation on O.O.D. effectiveness}}
Our first ablation experiment focused on evaluating the initial distillation process using the MS MARCO dataset with soft labels. To accomplish this, we generated logits with monoT5-3B and trained both T5-base and T5-small models for 10 epochs. As shown in Table~\ref{table:ablation}, rows 1-2 \& 5-6, both models demonstrated an improvement in their nDCG@10 scores compared to the baseline, which was trained using the hard labels from MS MARCO. Remarkably, the overall score increased in both scenarios, even though the models were not exposed to any BEIR passages during this phase.\footnote{The individual results for each dataset are presented in Appendix~\ref{appendix:full_results}.}

\paragraph{\textbf{Adding soft synthetic targets as a second distillation phase}} For the next experiment, we applied a second distillation step with synthetic soft labels on top of the model that we acquired from the last phase (monoT5 w/ soft human labels). For that, we used the 100K synthetic queries generated by InPars for each dataset indicated as ``in-domain'' and trained for 10 epochs. As shown in Table~\ref{table:ablation}, rows 3 \& 7, while it was expected that the in-domain datasets would have an increase in their nDCG@10 scores, we observe that the out-of-domain datasets also had improvements, suggesting that the model's generalization capabilities were enhanced.

\paragraph{\textbf{Using hard human targets for the first distillation phase}}
Finally, we investigated the impact of skipping the first phase of distillation on MS MARCO logits, and instead starting from a model that was trained on hard human labels (monoT5-small and monoT5-base) and directly training using the synthetic soft BEIR targets. As we can see in Table~\ref{table:ablation}, rows 3-4 \& 7-8, when comparing with the model that was trained using the soft human targets, the overall effectiveness was reduced. From this, we conclude that the distillation step that includes the soft human targets on MS MARCO is beneficial, as it improves the model's effectiveness in both in-domain and out-of-domain scenarios.

\paragraph{\textbf{Upper bound for soft distillation}}
To estimate the upper bound of the effectiveness that these models could attain through distillation, we repeated the process using \textit{real queries} from BEIR, (i.e., the validation queries) instead of the synthetic ones. Results presented in Table~\ref{table:ablation-upper} show that for both model sizes, there was an increase in effectiveness for the in-domain datasets, as the model was exposed to the evaluation queries during training. However, we also observed an increase in effectiveness for out-of-domain datasets, indicating that the synthetic queries used for training could be improved.

\begin{table}[ht]
\centering
\renewcommand{\arraystretch}{1.3}
\setlength{\tabcolsep}{2pt}
{%
\begin{tabularx}{\columnwidth}{@{}l *{4}{>{\centering\arraybackslash}X}@{}}
    \toprule
     & \multicolumn{2}{c}{\textbf{Sample Set 1}} & \multicolumn{2}{c}{\textbf{Sample Set 2}} \\ \cmidrule(lr){2-3} \cmidrule(lr){4-5}
    \multicolumn{1}{c}{{\textbf{Model}}} & In-domain & O.O.D. & In-domain & O.O.D. \\ \midrule
    InRanker-60M & 0.4768 & 0.4716 & 0.5558 & \textbf{0.3852} \\
    $\hookrightarrow$ w/ real queries & \textbf{0.4975} & \textbf{0.4719} & \textbf{0.5860} & 0.3813 \\ \midrule
    InRanker-220M & 0.4945 & 0.5028 & 0.5874 & 0.4083 \\
    $\hookrightarrow$ w/ real queries & \textbf{0.5242} & \textbf{0.5175} & \textbf{0.6159} & \textbf{0.4202} \\ \bottomrule
\end{tabularx}%
}
\vspace{0.1cm}
\caption{Upper bound effectiveness (nDCG@10) using real queries from BEIR for the distillation datasets. Bold indicates the best between using synthetic and real queries.}
\label{table:ablation-upper}
\end{table}

\section{Conclusion}
This paper introduces a method for distilling the knowledge of information retrieval models and improve upon previous work how to better use synthetic data, aimed at improving the out-of-domain effectiveness of students. The study reveals that, through this knowledge distillation process, smaller models can achieve results comparable to the teacher. This approach is particularly significant for applications where computational resources are limited or in production environments. The methodology involves two steps of distillation: (1) using a human-curated corpus, and (2) using synthetic data generated by an LLM. Consequently, the paper shows that it is possible to improve a reranker's capabilities in specific domains without requiring additional human-annotated labels. Finally, we observe that synthetic query generation could be improved since the real queries achieved a better out-of-domain effectiveness compared to the model trained solely on synthetic ones.

\bibliographystyle{ACM-Reference-Format}
\bibliography{refs}

\clearpage
\appendix

\section{Training Details}
\label{sec:training}
Table \ref{table:training} presents the parameters used for training the models using an A100 GPU with 80GB of VRAM. All experiments were conducted using the same learning rate of 7e-5 and the AdamW optimizer with its default hyperparameters in HuggingFace. The batch size was set to 32. For the 3B model, we used gradient checkpointing and gradient accumulation (to achieve an effective batch size of $2 \times 16$) due to memory constraints. During the generation of soft labels using the teacher model, we sampled $9$ non-relevant passages passage for each relevant passage, leading to 10 pairs of logits per query. Differently from InPars and Promptagator, which train a separate model for each dataset, InRanker is a single model trained on all 16 datasets from BEIR.

\begin{table}[ht]
\centering
\renewcommand{\arraystretch}{1.3}
\setlength{\tabcolsep}{4pt}
{%
\begin{tabular}{@{}ccccc@{}}
\toprule
\textbf{Parameters} &
  \textbf{Dataset} &
  \textbf{Steps} &
  \textbf{Epochs} &
  \textbf{\begin{tabular}[c]{@{}c@{}}Training \\ Duration\end{tabular}} \\ \midrule
\multirow{2}{*}{60M} & Human Soft & 400k & 10 & 7h    \\
 & Synthetic Soft & 329k & 1  & 5:30h \\ \midrule
\multirow{2}{*}{220M}  & Human Soft  & 400k & 10 & 15h   \\
 & Synthetic Soft  & 329k & 1  & 12h   \\ \midrule
\multirow{2}{*}{3B}    & Human Soft  & 400k & 10 & 300h  \\
 & Synthetic Soft     & 329k & 1  & 250h  \\ \bottomrule
\end{tabular}%
}
\vspace{0.1cm}
\caption{Training hyperparameters and duration using an A100-80GB GPU.}
\label{table:training}
\end{table}

\section{Datasets used in ablations}
\label{sec:datasets}

Table~\ref{table:composition_v1_v2} shows the datasets that were \textit{randomly} chosen for inclusion in each sample set, resulting in the use of 12 out of the 16 BEIR datasets (as some were not used for training at all).

\begin{table}[hb]
\renewcommand{\arraystretch}{1.1}
\setlength{\tabcolsep}{3pt}
\centering
\begin{tabular}{@{}ccc@{}}
\toprule
\textbf{Dataset} & \textbf{Sample Set 1} & \textbf{Sample Set 2} \\ \midrule
TREC-COVID &  & $\checkmark$ \\
NFCorpus & $\checkmark$ &  \\
BioASQ &  & $\checkmark$ \\
NQ & $\checkmark$ & $\checkmark$ \\
HotpotQA & $\checkmark$ & $\checkmark$ \\
Climate-FEVER &  &  \\
DBPedia & $\checkmark$ &  \\
TREC-NEWS &  &  \\
Robust04 &  & $\checkmark$ \\
ArguAna &  &  \\
Touché-2020 &  &  \\
Quora & $\checkmark$ &  \\
SCIDOCS & $\checkmark$ & $\checkmark$ \\
SciFact &  & $\checkmark$ \\
FiQA-2018 & $\checkmark$ & $\checkmark$ \\
Signal-1M & $\checkmark$ &  \\ \bottomrule
\end{tabular}
\vspace{0.1cm}
\caption{Composition of the two sample sets used in the ablation experiments, using datasets from the BEIR benchmark.}
\label{table:composition_v1_v2}
\end{table}

\section{Loss Function Ablation}
\label{sec:losses}
We tested different loss functions, including the KL divergence and MSE, to match the logits of the two models. Table~\ref{table:losses} shows the results, indicating that KL divergence was slightly worse for T5-small and that using only the true label in MSE as opposed to using both true and false labels also reduced the effectiveness. 

\begin{table}[hb]
\centering
{%
\begin{tabular}{@{}ccc@{}}
\toprule
\textbf{Parameters} & \textbf{Loss} & \textbf{Results} \\ \midrule
\multirow{3}{*}{60M} & MSE with normalized logits & 0.4807 \\
 & MSE with ``true'' logit only & 0.4748 \\ 
 & KL divergence & 0.4712 \\ \midrule
\multirow{2}{*}{220M} & MSE with normalized logits & 0.5008 \\
 & KL divergence & 0.5012 \\ \bottomrule
\end{tabular}%
}
\vspace{0.1cm}
\caption{Average nDCG@10 on 16 datasets of the BEIR benchmark with varying loss functions.}
\label{table:losses}
\end{table}

\section{Complete Results on BEIR}
\label{appendix:full_results}
Table~\ref{table:complete_results} presents the results obtained after distilling the models using soft labels from MS MARCO and BEIR. We can observe the impact of both proposed distillation steps, namely using soft human labels and soft synthetic labels, which bring significant effectiveness improvements over the base models. In particular, using logits from MS MARCO leads to an average of a 2-point nDCG@10 improvement for each model, while the subsequent finetuning phase with the synthetic BEIR queries further enhances their effectiveness by 4.5 points for T5-small and approximately 1.4 points for T5-base.

\begin{table*}[ht]
    \centering
    \renewcommand{\arraystretch}{1.2}
    \setlength{\tabcolsep}{3pt}
    {%
    \begin{tabular}{l|ccc|ccc|c}
    \toprule
    \multicolumn{1}{c|}{\textbf{Dataset}}  & \multicolumn{3}{c|}{\textbf{T5-small} (60M)}                     & \multicolumn{3}{c|}{\textbf{T5-base} (220M)}                             & \textbf{T5-3B} \\ \cline{2-8} 
               & \multicolumn{1}{c|}{Baseline} & 1st Step & + 2nd Step      & \multicolumn{1}{c|}{Baseline} & 1st Step        & + 2nd Step      & Teacher        \\ \hline
    TREC-COVID & \multicolumn{1}{c|}{0.6928}   & 0.7247   & \textbf{0.7775} & \multicolumn{1}{c|}{0.7775}   & 0.7643          & \textbf{0.7984} & 0.7936         \\
    NFCorpus   & \multicolumn{1}{c|}{0.3180}   & 0.3475   & \textbf{0.3547} & \multicolumn{1}{c|}{0.3570}   & 0.3639          & \textbf{0.3658} & 0.3801         \\
    BioASQ     & \multicolumn{1}{c|}{0.4880}   & 0.4648   & \textbf{0.5516} & \multicolumn{1}{c|}{0.5240}   & 0.5281          & \textbf{0.5652} & 0.5652         \\
    NQ         & \multicolumn{1}{c|}{0.4733}   & 0.5214   & \textbf{0.5469} & \multicolumn{1}{c|}{0.5674}   & 0.5855          & \textbf{0.5971} & 0.6251         \\
    HotpotQA   & \multicolumn{1}{c|}{0.5996}   & 0.6842   & \textbf{0.7563} & \multicolumn{1}{c|}{0.6950}   & 0.7546          & \textbf{0.7742} & 0.7595         \\
    Climate-FEVER    & \multicolumn{1}{c|}{0.2116} & 0.2488 & \textbf{0.2729} & \multicolumn{1}{c|}{0.2451} & 0.2739          & \textbf{0.2914} & 0.2835          \\
    DBPedia    & \multicolumn{1}{c|}{0.3437}   & 0.3745   & \textbf{0.4451} & \multicolumn{1}{c|}{0.4195}   & 0.4446          & \textbf{0.4650} & 0.4719         \\
    TREC-NEWS  & \multicolumn{1}{c|}{0.3848}   & 0.4478   & \textbf{0.4646} & \multicolumn{1}{c|}{0.4475}   & \textbf{0.4808} & 0.4695          & 0.4806         \\
    Robust04   & \multicolumn{1}{c|}{0.4222}   & 0.4782   & \textbf{0.5386} & \multicolumn{1}{c|}{0.5016}   & 0.5588          & \textbf{0.5774} & 0.6171         \\
    ArguAna    & \multicolumn{1}{c|}{0.1274}   & 0.1098   & \textbf{0.2466} & \multicolumn{1}{c|}{0.1946}   & 0.2431          & \textbf{0.2873} & 0.3824         \\
    Touché-2020      & \multicolumn{1}{c|}{0.2643} & 0.2557 & \textbf{0.2883} & \multicolumn{1}{c|}{0.2773} & \textbf{0.2991} & 0.2897          & 0.3026          \\
    Quora      & \multicolumn{1}{c|}{0.8259}   & 0.8246   & \textbf{0.8335} & \multicolumn{1}{c|}{0.8230}   & 0.8418          & \textbf{0.8427} & 0.8347         \\
    SCIDOCS    & \multicolumn{1}{c|}{0.1436}   & 0.1526   & \textbf{0.1788} & \multicolumn{1}{c|}{0.1649}   & 0.1746          & \textbf{0.1911} & 0.1978         \\
    SciFact    & \multicolumn{1}{c|}{0.6963}   & 0.7022   & \textbf{0.7490} & \multicolumn{1}{c|}{0.7356}   & 0.7505          & \textbf{0.7618} & 0.7773         \\
    FiQA-2018  & \multicolumn{1}{c|}{0.3377}   & 0.3712   & \textbf{0.4043} & \multicolumn{1}{c|}{0.4136}   & 0.4374          & \textbf{0.4431} & 0.5068         \\
    Signal-1M  & \multicolumn{1}{c|}{0.2711}   & 0.2612   & \textbf{0.2820} & \multicolumn{1}{c|}{0.2771}   & 0.2910          & \textbf{0.2926} & 0.3004         \\ \hline
    Average & \multicolumn{1}{c|}{0.4125} & 0.4356 & \textbf{0.4807} & \multicolumn{1}{c|}{0.4638} & 0.4870          & \textbf{0.5008} & \textbf{0.5174} \\ \bottomrule
    \end{tabular}%
    }
    \vspace{0.1cm}
    \caption{nDCG@10 values for each dataset after two steps of distillation.}
    \label{table:complete_results}
    \end{table*}

\end{document}